\documentclass[a4paper,11pt]{article}
\usepackage{pos}

\usepackage{subcaption}

\usepackage{mathtools}
\usepackage{physics}

\usepackage[disable]{todonotes}

\makeatletter
\g@addto@macro\bfseries{\boldmath}
\makeatother

\newcommand{\m}[1]{\mathrm{#1}}

\let\phi\varphi
\let\epsilon\varepsilon

\title{Distillation and position-space sampling for local multiquark interpolators}

\author*[a]{Andres~Stump}
\author[b]{Jeremy~R.~Green}

\affiliation[a]{Institut für Physik, Humboldt-Universität zu Berlin,\\
  Zum Großen Windkanal 2, 12489 Berlin, Germany}

\affiliation[b]{Deutsches Elektronen-Synchrotron DESY,\\
  Platanenallee 6, 15738 Zeuthen, Germany}

\emailAdd{andres.stump@hu-berlin.de}
\emailAdd{jeremy.green@desy.de}

\abstract{Distillation in lattice QCD is a smearing method that uses the lowest eigenvectors of the spatial Laplacian to construct a subspace in which the Dirac operator can be fully inverted. However, local multiquark interpolators are expensive in this framework because the cost of the contractions scales with a high power of the number of Laplacian eigenvectors. To address this, a position-space sampling method within distillation is presented that avoids this cost scaling. Our simulations show that this method works well for meson operators, but also for local tetraquark operators. In a preliminary study, we investigate the relevance of the latter for the ground state energy of the $T_{cc}(3875)^+$ tetraquark. There we find a downward shift in the lowest level when local tetraquark operators are added to a basis of bilocal scattering operators in the variational method. However, this shift is small compared to the error of the energy.}

\FullConference{The 41st International Symposium on Lattice Field Theory (LATTICE2024)\\
 28 July - 3 August 2024\\
Liverpool, UK\\}

\begin{document}
\maketitle

\section{Introduction}

Many exotic hadrons have been discovered in the last decade, including the $T_{cc}(3875)^+$ tetraquark observed and studied by the LHCb collaboration~\cite{LHCb:2021vvq, LHCb:2021auc} but also the $d^*(2380)$ dibaryon observed by the WASA-at-COSY collaboration~\cite{PhysRevLett.106.242302}. The preferred way to study exotic hadrons using lattice QCD is Lüscher's finite-volume quantization conditions, which require finite-volume energies. The standard way to obtain these energies is the variational method, which requires using many operators carrying the quantum numbers of the state of interest. Since the inner structure of these exotic hadrons is generally unknown, it is desirable to use a variety of operators with different spin, color and spatial structures to arrive at the correct low-lying spectrum. As bound states or resonances, exotic hadrons are in some sense local, so one wants to include local operators in addition to bilocal scattering ones. A good framework for studying bilocal scattering operators is distillation~\cite{HadronSpectrum:2009krc}, and it has been used successfully for tetraquarks, nucleon-nucleon scattering and dibaryons~\cite{Padmanath:2022cvl, PhysRevLett.127.242003}. Although distillation has also been used for local multiquark operators~\cite{Cheung:2017tnt, Ortiz-Pacheco:2023ble}, it is computationally expensive due to the high-rank tensors that appear. In this work, we present a position-space sampling method that makes local multiquark operators more affordable within distillation.\footnote{Recently, another approach was proposed to solve this problem~\cite{Lang:2024syy}.}

\section{Distillation Method}

The distillation method described in~\cite{HadronSpectrum:2009krc, Morningstar:2011ka} uses Laplacian Heaviside smearing for the quarks. This is done by replacing the quark field $\psi(t)$ by the spatially smeared field ${\psi_\m{sm}(t)=\mathcal{S}(t)\,\psi(t)}$ where the smearing kernel $\mathcal{S}(t) = V(t)V(t)^\dag$ is defined by the matrix
\begin{equation}
  V(t) = \left(v^{(1)}(t), v^{(2)}(t), \dots, v^{(N)}(t)\right)
\end{equation}
which contains eigenvectors of the gauge covariant Laplace operator $\Delta(t)$ on the lattice. More precisely, $V(t)$ contains those $N$ eigenvectors $v^{(k)}(t)$ of $-\Delta(t)$ with the smallest eigenvalues. The Laplacian and thus its eigenvectors have a time dependence since they are defined in terms of the gauge links on a specific time slice. We will also refer to the $v^{(k)}(t)$ as Laplace modes and call $k$ the Laplace mode index.

The Laplacian acts on a $3\abs{\Lambda_3}$-dimensional color and position space (we use $\Lambda_3$ to denote the spatial lattice) so $\mathcal{S}$ is a projector from this large space to an $N$-dimensional subspace spanned by the lowest $N$ eigenvectors of $-\Delta(t)$. Usually, $N$ is chosen to be of the order of a few tens or hundreds. In this work, we used $N=32$ throughout. Thus, the "distillation" space spanned by these eigenvectors is small enough that the Dirac operator $D_f$ (for flavour $f$) can be inverted exactly in this subspace. The $4N\times4N$ matrix obtained in this way is called a perambulator and is defined as
\begin{equation}
    (\tau_f)_{\alpha\beta}(t', t) = V(t')^\dag \cdot (D_f^{-1})_{\alpha\beta}(t', t) \cdot V(t).
\end{equation}
Here we have explicitly written out the spinor indices, but suppressed the Laplace mode, color and position indices. To compute this matrix we need four inversions of the Dirac operator on each Laplace mode at time $t$. Then we project the solutions onto the Laplace modes at time $t'$. Finally, the smeared propagator is given by
\begin{equation} \label{eq:smeared_propagator}
  (D_{\m{sm}, f}^{-1})_{\alpha\beta}(t', t) = V(t') \cdot (\tau_f)_{\alpha\beta}(t', t) \cdot V(t)^\dag
\end{equation}
in terms of the perambulator and Laplace modes.

To demonstrate how the distillation method works we consider the simple case of a charged pion using the operator
\begin{equation} \label{eq:pion_operator}
  \mathcal{O}^\pi(t) = \sum_{\vb*{x} \in \Lambda_3} e^{-i\vb*{p}\cdot\vb*{x}} (\overline{d} \gamma_5 u)(\vb*{x}, t)
\end{equation}
with momentum $\vb*{p}$.
The resulting two-point function is then given by
\begin{equation} \label{eq:pion_two_point_function}
  C(t', t)
    = \sum_{\vb*{x}', \vb*{x} \in \Lambda_3} e^{-i\vb*{p}\cdot(\vb*{x'}-\vb*{x})}
      \ev{\tr\left[
        \gamma_5D^{-1}_u(\vb*{x'}, t'; \vb*{x}, t)\cdot
        \gamma_5 D^{-1}_d(\vb*{x}, t; \vb*{x}', t')
  \right]}_G
\end{equation}
where $\ev{\dots}_G$ is the expectation value over the gauge fields only. Plugging in the expression for the smeared propagator \eqref{eq:smeared_propagator} we get the smeared correlator
\begin{equation} \label{eq:smeared_pion_two_point_function}
    C_\m{sm}(t', t) = \ev{\tr\left[ 
            \Phi(t')\cdot \gamma_5 \tau_u(t', t)\cdot
            \Phi(t)^\dag\cdot \gamma_5 \tau_d(t, t')
        \right]}_G \quad
\end{equation}
where we have collected the momentum projection and the eigenvectors in the so-called mode doublets $\Phi(t)$ defined as
\begin{equation} \label{eq:mode_doublet}
  \Phi^{(k, l)}(t) = \sum_{\vb*{x} \in \Lambda_3} e^{-i\vb*{p}\cdot\vb*{x}} \, v_a^{(k)}(\vb*{x}, t)^*  \,v_a^{(l)}(\vb*{x}, t)
\end{equation}
which form an $N\times N$ matrix. Here we have explicitly written out the position index $\vb*{x}$ and the color index $a$. Using this matrix for the contractions is more practical than recomputing or storing the Laplacian eigenvectors.

Next, we will look at more complicated operators where we will explain the difficulties that arise for local multiquark operators. Before we do so, let us briefly address the question of how we should scale the number of eigenvectors $N$ for different lattice spacings and volumes. In~\cite{Morningstar:2011ka} it was shown that to keep the physical smearing radius constant we have to scale $N$ proportional to the physical volume $V = a^3|\Lambda_3|$ ($a$ is the lattice spacing). Thus, larger physical volumes result in a bigger computational cost for calculating the perambulators, but also the contractions become more expensive which poses a problem for local multiquark interpolators which we will discuss next.

\subsection{Local multiquark operators in distillation}

Like for mesons, distillation can also be used for baryons~\cite{PhysRevLett.127.242003}. But where for mesons we introduced the mode doublets $\Phi^{(k, l)}(t)$, for baryon correlators we have to introduce a different tensor to incorporate the Laplace modes and the momentum projection: the so-called mode triplets. They are defined by
\begin{equation}
  \Phi^{(k, l, m)}(t) = \sum_{\vb*{x} \in \Lambda_3} e^{-i\vb*{p}\cdot\vb*{x}} \, 
      \epsilon_{abc} \, v_a^{(k)}(\vb*{x}, t) v_b^{(l)}(\vb*{x}, t) v_c^{(m)}(\vb*{x}, t).
\end{equation}
This is a rank 3 tensor because we get one Laplace mode index for each quark in the baryon operator. Similarly, two-point functions of bilocal meson-meson and baryon-baryon scattering operators can be written as tensor contractions of perambulators with mode doublets and triplets respectively. This means that for mesons and meson scattering the computational cost of the contractions scales as $N^3$, since they consist of matrix-matrix multiplications. For baryons and baryon scattering, it scales as $N^4$ because we are contracting matrices with rank 3 tensors.

It is more difficult when we consider local multiquark operators. For the $T_{cc}(3875)^+$ tetraquark and the $d^*(2380)$ dibaryon, these operators are of the form (spin and color structure suppressed)
\begin{equation}
  \mathcal{O}^T(t) \sim \sum_{\vb*{x} \in \Lambda_3}
  e^{-i\vb*{p}\cdot\vb*{x}} (cc\bar{u}\bar{d})(\vb*{x}, t)
  \quad \text{and} \quad
  \mathcal{O}^H(t) \sim \sum_{\vb*{x} \in \Lambda_3} 
        e^{-i\vb*{p}\cdot\vb*{x}} (uuuddd)(\vb*{x}, t).
\end{equation}
In these cases, the mode doublets and triplets have to be replaced by rank 4 and rank 6 tensors respectively. As a result, the cost of the contractions scales as $N^5$ and $N^7$. This is very expensive, even for a relatively small $N$, such as $N=32$, as used in our case. In addition, as mentioned earlier, to have a constant smearing radius we have to scale $N$ proportional to the physical volume. This makes local tetraquark and hexaquark operators prohibitively expensive in large volumes.

The problem here is that by summing over $\vb*{x}$ first and then over the Laplace mode indices, we construct high-rank tensors which make the contractions expensive. However, summing over the sink and source positions last, as in \eqref{eq:pion_two_point_function}, results in a computational cost proportional to $N\abs{\Lambda_3}^2$ for computing the smeared propagator $D^{-1}_\m{sm}$ (cf. \eqref{eq:smeared_propagator})\footnote{Computing $D^{-1}_\m{sm}$ also involves a contraction that scales as $N^2\abs{\Lambda_3}$ (similar volume scaling as $N\abs{\Lambda_3}^2$), but $N \ll \abs{\Lambda_3}$.}
and $\abs{\Lambda_3}^2$ for the contractions. This has a better volume scaling but is still too expensive. The solution we propose here is to change the summation order, but then make computing $D^{-1}_\m{sm}$ and the contractions affordable by using a position-space sampling method. This will be discussed in the next section.

\section{Position-space sampling}

In this section, we explain the position-space sampling method we propose for local multiquark operators within distillation. The idea is that we compute the smeared propagator \eqref{eq:smeared_propagator} only on random subspaces $\tilde{\Lambda}_3 \subset \Lambda_3$ instead of on the full spatial lattice $\Lambda_3$. More precisely we choose a random subspace $\tilde{\Lambda}_3'$ for the sink and another one $\tilde{\Lambda}_3$ for the source and compute ${D_{\m{sm}, f}^{-1}(\vb*{x}', t'; \vb*{x}, t)}$ for all $\vb*{x}' \in \tilde{\Lambda}_3'$ and $\vb*{x} \in \tilde{\Lambda}_3$. For this, we need the perambulator and the Laplace modes on these two subspaces. To compute the correlator, we then only sum over $\tilde{\Lambda}_3'$ and $\tilde{\Lambda}_3$. For the charged pion (cf. \eqref{eq:pion_two_point_function}) the two-point function in position-space sampling is
\begin{equation} \label{eq:pos_space_sampling_pion}
    C_\m{sm}(t', t) = \Big<
        \frac{|\Lambda_3|^2}{|\tilde{\Lambda}'_3||\tilde{\Lambda}_3|}\,
        \sum_{
        \mathclap{\substack{\vb*{x}' \in \tilde{\Lambda}'_3 \\
                  \vb*{x} \in \tilde{\Lambda}_3}}
      }
      e^{-i\vb*{p}\cdot(\vb*{x}' - \vb*{x})} 
        \tr\left[
            \gamma_5 D_{u,\,\m{sm}}^{-1}(\vb*{x}',t'; \vb*{x},t) \cdot
            \gamma_5 D_{d,\,\m{sm}}^{-1}(\vb*{x},t; \vb*{x}', t')
        \right]\Big>_{G, \tilde{\Lambda}_3}
\end{equation}
where we have added the prefactor ${|\Lambda_3|^2/(|\tilde{\Lambda}'_3||\tilde{\Lambda}_3|)}$ to ensure the correct normalization. The expectation value $\ev{\dots}_{G, \tilde{\Lambda}_3}$ is now an expectation value both over the gauge fields and over the two random subspaces. Thus, by using position-space sampling, we get a stochastic estimator for the correlator that is different from the one in \eqref{eq:smeared_pion_two_point_function}.

Two natural choices for these subspaces are fully random subspaces of $\Lambda_3$ and randomly displaced sparse grids. We chose the latter. More specifically, we used subspaces of the form
\begin{equation}
  \tilde{\Lambda}_3 = \{a \vb*{n} + \vb*{\tilde{x}}\, | n_k = 0,\, N_\m{sep},\, 2N_\m{sep},\, \dots,\, N_s-N_\m{sep} \quad (k = 1, 2, 3) \}
\end{equation}
with a point separation $N_\m{sep}$ which divides the spatial extent $N_s$ of the lattice (in lattice units) and with a random offset $\vb*{\tilde{x}} \in \Lambda_3$.
These are similar to the sparse grids used in~\cite{Detmold:2019fbk, Li:2020hbj}, but the random displacement that we added ensures that we sample from the full spatial lattice. We chose to sample $\vb*{\tilde{x}}$ randomly for each gauge configuration and source time $t$, and also separately at the sink and the source. By splitting the expectation value as $\ev{\dots}_{G, \tilde{\Lambda}_3} = \ev{\ev{\dots}_G}_{\tilde{\Lambda}_3}$ one can show that using these random sparse grids we get an estimator for the correlator that is unbiased. Thus, randomly displacing the sparse grids ensures a full momentum projection. Consequently, the point separation $N_\m{sep}$ only affects the variance of this estimator, which can be reduced by decreasing $N_\m{sep}$. In addition, keeping the offsets the same for all sink times $t'$ preserves correlations between different $t'$.

Using this method we avoid the cost scaling of the contractions with a high power of the number of Laplace modes $N$, and we reduce the $N\abs{\Lambda_3}^2 = NN_s^6$ scaling down to an $N\abs{\tilde{\Lambda}_3}^2 = N(N_s/N_\m{sep})^6$ scaling. If we increase the physical volume $V$, we want to keep $aN_\m{sep}$ constant which results in a $V^3$ scaling of the computational cost. This is a huge improvement compared to the $V^5$ or $V^7$ scaling for local tetraquark or hexaquark operators without position-space sampling. So this method makes local multiquark interpolators possible in larger volumes. However, we still need to investigate which value to choose for the point separation $N_\m{sep}$. This will be addressed in the next section.

\section{Results for meson and tetraquark operators}

In this section, we will show finite-volume spectroscopy results for meson and tetraquark operators using the position-space sampling method that we have introduced. We will study the variance of the position-space sampling estimator \eqref{eq:pos_space_sampling_pion} as a function of the point separation $N_\m{sep}$ in the sparse grids. In a preliminary study, we will also investigate the influence of local tetraquark operators on the ground state energy of the $T_{cc}$ tetraquark.  All errors in this section only include the statistical error that we computed using the $\Gamma$-method~\cite{WOLFF2004143}.

\subsection{Lattice setup}

The simulation was performed on the B450 CLS~\cite{Bruno:2014jqa} gauge ensemble using $O(a)$-improved Wilson fermions at the $SU(3)$ flavour symmetric point with $m_\pi = m_K \approx 417$~MeV. This is a $64\times32^3$ lattice with $\beta = 3.46$ which corresponds to a lattice spacing of $a = 0.0762$~fm. For the valence charm quarks, we used the same fermion action tuned such that the $D$ meson mass matches the average of the physical $D^0$, $D^+$ and $D_s$ meson masses. We used all 1612 configurations with 8 source times each. The number of Laplace modes was always $N = 32$.

\subsection{Dependence on $N_\m{sep}$: single-meson operators}

\begin{figure}
  \begin{subfigure}{0.5\textwidth}
    \includegraphics[width=\textwidth]{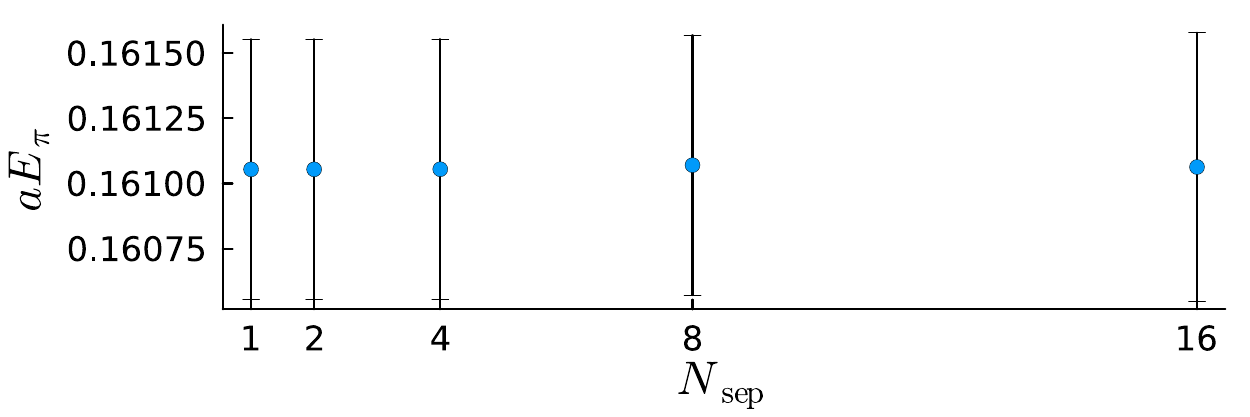}
  \end{subfigure}
  \hfill
  \begin{subfigure}{0.5\textwidth}
    \includegraphics[width=\textwidth]{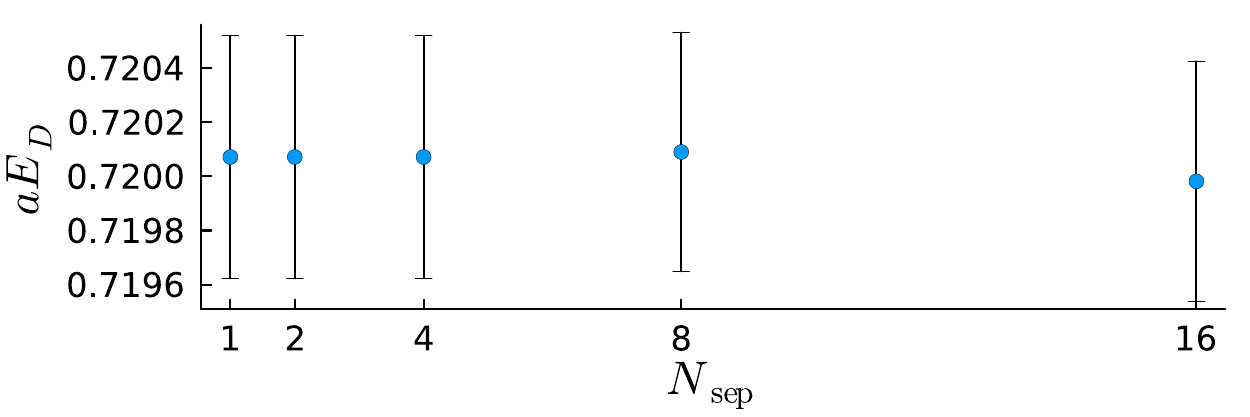}
  \end{subfigure}
  \caption{Pion (left) and $D$ meson (right) energy for zero momentum. They are plotted as a function of the point separation $N_\m{sep}$ in the sparse grids used for position-space sampling.}
  \label{fig:pion_D_energy_vs_Nsep}
\end{figure}

First, we applied the position-space sampling approach to the charged pion operator \eqref{eq:pion_operator}. We computed the pion two-point function for different point separations $N_\m{sep}$ and extracted the energy from a plateau fit to the effective mass. To isolate the dependence on $N_\m{sep}$, we used the same plateau range for all point separations. The resulting energies for zero momentum are shown as a function of $N_\m{sep}$ on the left of Figure \ref{fig:pion_D_energy_vs_Nsep}.
The point where $N_\m{sep} = 1$ corresponds to using the full spatial lattice, i.e. no position-space sampling, and increasing values of $N_\m{sep}$ correspond to increasingly sparse grids. There is no increase in error compared to the $N_\m{sep}=1$ value, even when going to $N_\m{sep} = 16$, which corresponds to using only 2 points in each spatial direction. We repeated this calculation for momentum $\vb*{p} = \frac{2\pi}{L}(1, 0, 0)$ and also found no increase in error up to $N_\m{sep}=8$, but a small increase for $N_\m{sep}=16$. In all cases, the energy was consistent with the $N_\m{sep}=1$ value, as expected from an unbiased estimator.

Next, we repeated the same analysis for the $D$ meson using
${\mathcal{O}^D(t) = \sum_{\vb*{x} \in \Lambda_3} e^{-i\vb*{p}\cdot\vb*{x}} (\overline{u} \gamma_5 c)(\vb*{x}, t)}$
as the interpolating operator.
The results are shown on the right of Figure \ref{fig:pion_D_energy_vs_Nsep}. Like for the pion, we see no increase in error up to $N_\m{sep}=8$, but we see a slight increase for $N_\m{sep}=16$. For momentum $\vb*{p} = \frac{2\pi}{L}(1, 0, 0)$, we also found no increase in error up to $N_\m{sep}=8$, but a bigger increase for $N_\m{sep}=16$. The energy was always consistent with the $N_\m{sep}=1$ value. We obtained similar results for the $D^*$ meson using the operator ${\mathcal{O}^{D^*}(t) = \sum_{\vb*{x} \in \Lambda_3} e^{-i\vb*{p}\vb*{x}}(\overline{u} \gamma_i c) (\vb*{x}, t)}$.

Our conclusion from these results is that the statistical error from position-space sampling for these two-point functions is negligible compared to the Monte Carlo error starting at $N_\m{sep}=8$. As a consequence, using a smaller point separation $N_\m{sep}$ doesn't reduce the error any further.

\subsection{Dependence on $N_\m{sep}$: tetraquark operators}

\begin{figure}
  \begin{subfigure}{0.5\textwidth}
    \includegraphics[width=\textwidth]{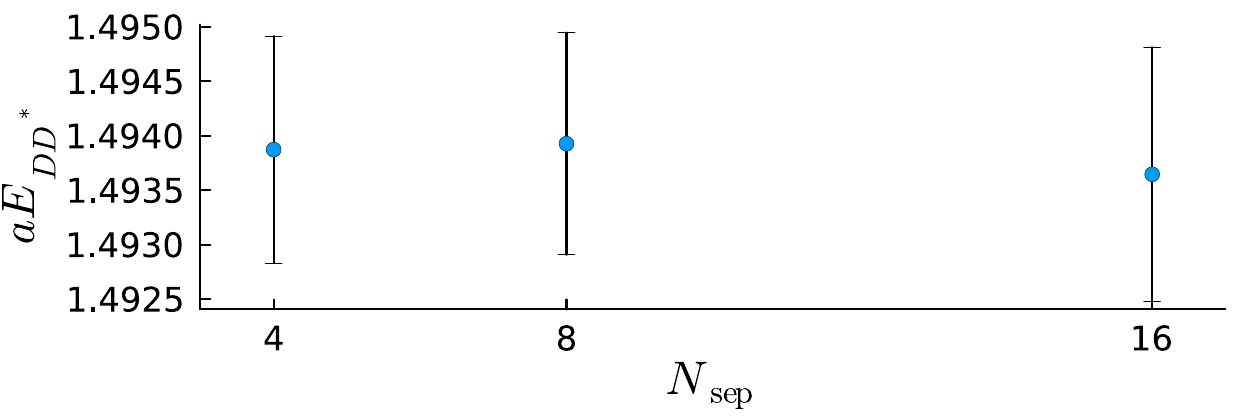}
  \end{subfigure}
  \hfill
  \begin{subfigure}{0.5\textwidth}
    \includegraphics[width=\textwidth]{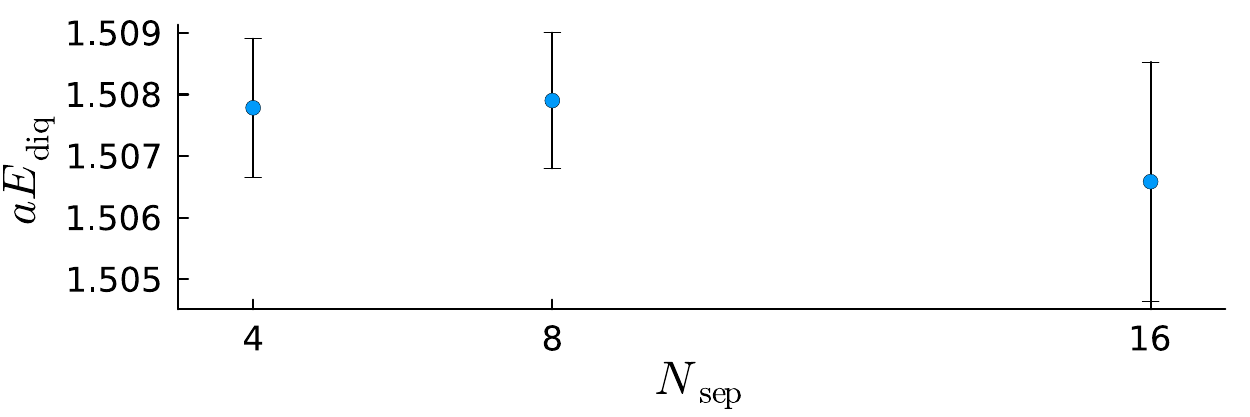}
  \end{subfigure}
  \caption{Plateau values obtained from a constant fit to the effective masses of local $DD^*$ (left) and diquark-antidiquark (right) two-point functions with zero momentum. They are plotted as a function of the point separation $N_\m{sep}$ in the sparse grids used for position-space sampling.}
  \label{fig:tetraquark_energy_vs_Nsep}
\end{figure}

Next, we applied the position-space sampling method to tetraquark operators relevant for the $T_{cc}(3875)^+$ which is a $I(J^P)=0(1^+)$ state with minimal quark content $cc\bar{u}\bar{d}$. For this, we used the local $DD^*$ operator
\begin{equation} \label{eq:operator_DDstar_local}
  \mathcal{O}_\m{local}^{DD^*}(t) = \sum_{\vb*{x} \in \Lambda_3} e^{-i\vb*{p}\cdot\vb*{x}}
      (\overline{u} \gamma_5 c \; \overline{d} \gamma_i c)(\vb*{x}, t)
      - \{u \leftrightarrow d\}
\end{equation}
as well as the local diquark-antidiquark operator
\begin{equation} \label{eq:operator_diq_local}
	\mathcal{O}_\m{local}^\m{diq}(\vb*{p}, t) =
		\sum_{\vb*{x} \in \Lambda_3} e^{-i\vb*{p}\cdot\vb*{x}}
		(\epsilon_{abc} \, c_b^T \, C\gamma_i \, c_c \;\;
		\epsilon_{ade} \,\overline{u}_d \, C\gamma_5 \, \overline{d}_e^T)(\vb*{x}, t)
\end{equation}
which is in the $(\vb*{\overline{3}}_c \otimes \vb*{3}_c)_{\vb*{1}_c}$ color representation. Here $C$ is the charge conjugation matrix and the Roman letters are color indices. For these two operators, we only used zero momentum, so they belong to the rest frame $T_1^+$ irreducible representation of the octahedral group.  Like for the quark-antiquark mesons, we computed the two-point functions for these operators for different point separations $N_\m{sep}$. To obtain plateau values whose errors we can compare, we fitted constants to the effective masses at late times. However, these fit results are not expected to be the correct finite-volume energy, since these operators are not optimized and the multiparticle energies are closely spaced. To actually extract the ground state energy, we later applied the variational method. In Figure \ref{fig:tetraquark_energy_vs_Nsep} we show the plateau values resulting from these fits.
Since the computational cost for computing these two-point functions scales as ${N_\m{sep}}^{-6}$, and since the contractions are more expensive than for single mesons, the smallest point separation we used is $N_\m{sep}=4$. However, as with the meson operators, we see no reduction in error when going from $N_\m{sep}=8$ to $N_\m{sep}=4$ for either operator. So we conclude that also for these more complicated operators, the statistical error is dominated by the Monte Carlo error starting at $N_\m{sep}=8$. This is the value for $N_\m{sep}$ that we use for the remaining calculations.

\subsection{Influence of local operators on $T_{cc}$ spectrum}

The influence of local operators on the ground state energy of the $T_{cc}$ tetraquark has been discussed in the literature~\cite{Collins:2024sfi, vujmilovic2024tccplanewaveapproach}. To investigate this, we used two bilocal $DD^*$ scattering operators to extract the ground state energy of the $T_{cc}$ using the variational method. We then added the two local operators (\ref{eq:operator_DDstar_local}, \ref{eq:operator_diq_local}) to this operator basis to study their effect on the energy. The two bilocal operators that we used are
\begin{equation} \label{eq:operator_DDstar_bilocal}
  \mathcal{O}_\m{bilocal}^{DD^*}(t) =
      \sum_{\vb*{p}^2 = n\left(\frac{2\pi}{L}\right)^2}
      \sum_{\vb*{x}_1, \vb*{x}_2 \in \Lambda_3}
      e^{-i\vb*{p}\cdot(\vb*{x}_1 - \vb*{x}_2)}
        (\overline{u} \gamma_5 c)(\vb*{x}_1, t) \;
        (\overline{d} \gamma_i c)(\vb*{x}_2, t)
        - \{u \leftrightarrow d\}
\end{equation}
for $n=0, 1$. They are also in the rest frame $T_1^+$ irreducible representation. To compute the full correlator matrix of all four operators, we used the mode doublets for the bilocal operators and position-space sampling for the local operators. We then extracted the lowest energy levels using the generalized eigenvalue problem (GEVP) method described in~\cite{Blossier:2009kd}.

\begin{figure}
  \begin{subfigure}{0.5\textwidth}
    \includegraphics[width=1\textwidth]{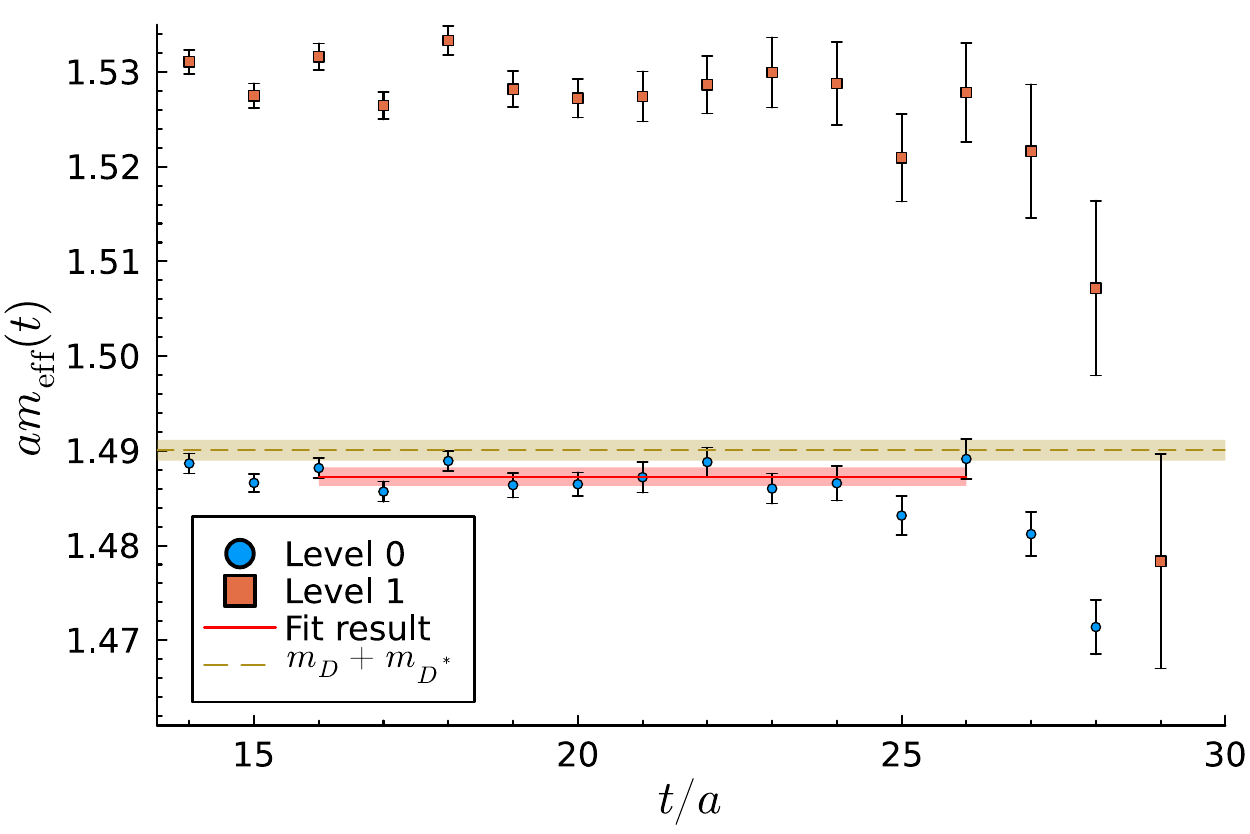}
  \end{subfigure}
  \hfill
  \begin{subfigure}{0.5\textwidth}
    \includegraphics[width=\textwidth]{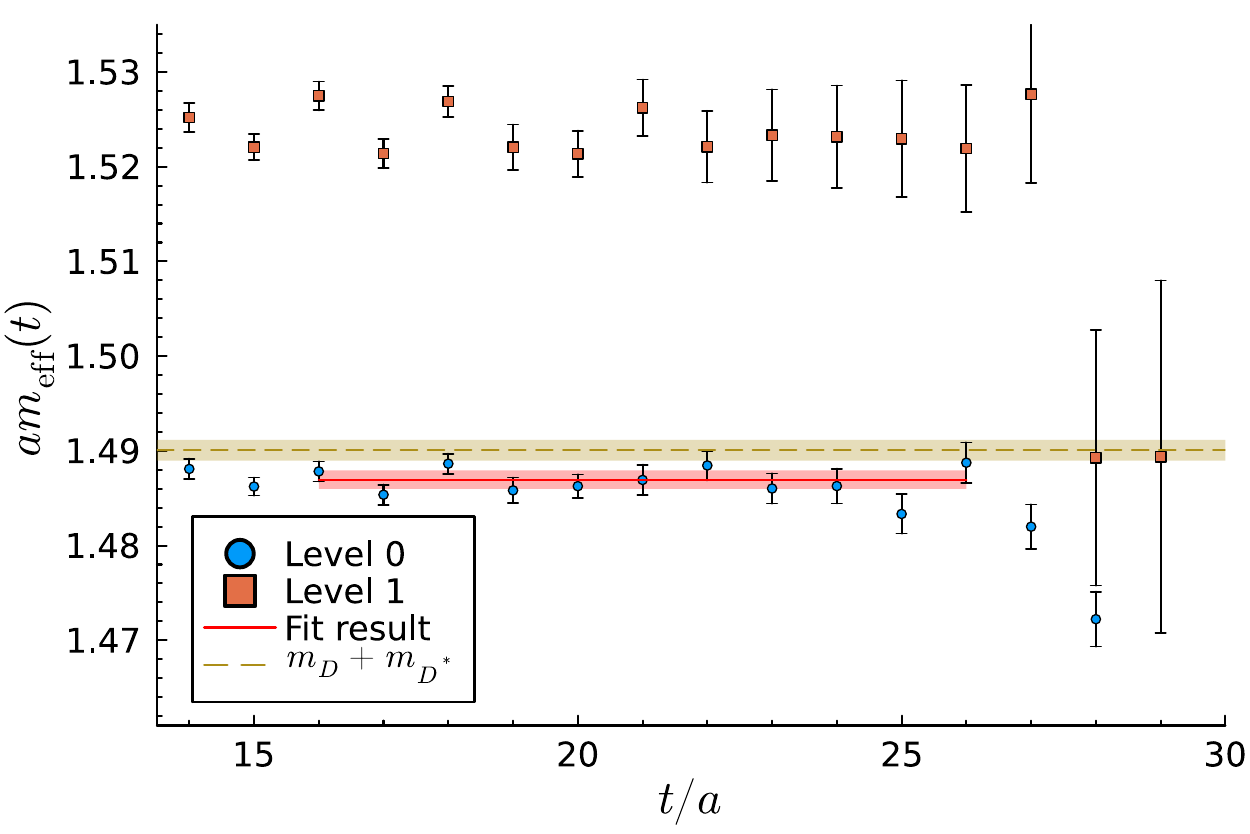}
  \end{subfigure}
  \caption{Effective masses of the two lowest energy levels of the $T_{cc}$ tetraquark. They were obtained using the variational method for the two bilocal operators (left) and when also including the two local operators (right). The red line shows the ground state energy extracted using a plateau fit and the orange dashed line shows the $m_D + m_{D^*}$ value for this ensemble.}
  \label{fig:tetraquark_effective_masses}
\end{figure}

In Figure \ref{fig:tetraquark_effective_masses} we show the effective masses of the two lowest energy levels when using only the two bilocal operators (left) and when including the two local operators (right).
We extracted the ground state energy using a plateau fit. As the energy of the $T_{cc}$ was measured to be slightly below the $DD^*$ mass threshold, we show the $m_D + m_{D^*}$ value that we computed on this ensemble as a reference. In the GEVP we set ${t_0/a = 13}$ such that the condition~\cite{Blossier:2009kd} ${t_0 \geq t/2}$ is satisfied over the whole plateau range we used for the fit. We see that the ground state energy decreases slightly when the local operators are included. The resulting value is $aE = 1.48696(98)$, and the difference is ${a\Delta E = 0.00032(3)}$ when taking the correlation into account. This is a significant decrease, but it is small compared to the error of the energy.

For this study, we only used two nonlocal operators, which is a small number. Since in the noninteracting limit, the first excited state is doubly degenerate, including further bilocal operators could result in a similar decrease. We need to investigate this before we can make a final statement about the influence of local operators on the ground state energy of the $T_{cc}$. In addition to this small shift in the lowest level, we see a larger shift in the first excited state, in agreement with~\cite{vujmilovic2024tccplanewaveapproach}. This will be studied in more detail when we include more bilocal operators.

\section{Conclusion}

We have presented a position-space sampling method for local multiquark operators within distillation that avoids a strong cost scaling in the physical volume. This is a new unbiased stochastic estimator for correlation functions which uses randomly displaced sparse grids instead of the full spatial lattice to perform the contractions. We have shown that this method works well for meson and local tetraquark operators: we can use a large point separation in the sparse grids while still achieving a statistical error that is dominated by the Monte Carlo error.

In a preliminary study, we have used this method to investigate the influence of local operators on the ground level of the $T_{cc}(3875)^+$. Adding two local operators to a basis of two bilocal scattering operators resulted in a downward shift in the finite-volume energy of the ground state. However, this shift is small compared to the error of the ground state energy and including more bilocal operators could result in a similar shift. Therefore, we plan to investigate the influence of local operators more thoroughly using further bilocal operators.

\acknowledgments

Calculations for this project used resources on the supercomputers JURECA~\cite{krause2018jureca} and JU\-WELS~\cite{krause2019juwels} at Jülich Supercomputing Centre (JSC). The raw distillation data were computed using QDP++~\cite{Edwards:2004sx}, PRIMME~\cite{PRIMME}, and the deflated SAP+GCR solver from openQCD~\cite{openQCD}. Contractions were performed using TensorOperations.jl~\cite{TensorOperations.jl}, the Monte Carlo analysis was done using ADerrors.jl~\cite{ADerrors.jl} and the plots were prepared with Plots.jl~\cite{christ2022plotsjluserextendable}. This work was funded by the  Deutsche Forschungsgemeinschaft (DFG, German Research Foundation) - Project number 417533893/GRK2575 "Rethinking Quantum Field Theory". We thank Renwick J. Hudspith for code development and are grateful to our colleagues within the CLS initiative for sharing ensembles.

\bibliographystyle{JHEP}
\bibliography{References}

\end{document}